# Development of a GPU-based Monte Carlo dose calculation code for coupled electron-photon transport

**Xun Jia[1], Xuejun Gu[1], Josep Sempau[2], Dongju Choi[3], Amitava Majumdar[3], and Steve B. Jiang[1]**

[1]Department of Radiation Oncology, University of California San Diego, La Jolla, CA 92037-0843, USA
[2]Institut de Techniques Energetiques, Universitat Politecnica de Catalunya, Diagonal 647, E-08028, Barcelona, Spain
[3]San Diego Supercomputer Center, University of California San Diego, La Jolla, CA 92093, USA

E-mail: sbjiang@ucsd.edu

Monte Carlo simulation is the most accurate method for absorbed dose calculations in radiotherapy. Its efficiency still requires improvement for routine clinical applications, especially for online adaptive radiotherapy. In this paper, we report our recent development on a GPU-based Monte Carlo dose calculation code for coupled electron-photon transport. We have implemented the Dose Planning Method (DPM) Monte Carlo dose calculation package (Sempau *et al*, *Phys. Med. Biol.*, 45(2000)2263-2291) on GPU architecture under CUDA platform. The implementation has been tested with respect to the original sequential DPM code on CPU in phantoms with water-lung-water or water-bone-water slab geometry. A 20 MeV mono-energetic electron point source or a 6 MV photon point source is used in our validation. The results demonstrate adequate accuracy of our GPU implementation for both electron and photon beams in radiotherapy energy range. Speed up factors of about 5.0 ~ 6.6 times have been observed, using an NVIDIA Tesla C1060 GPU card against a 2.27GHz Intel Xeon CPU processor.

## 1. Introduction

Dose calculation is of central importance in radiotherapy treatment planning. Among all algorithms developed for solving this problem, Monte Carlo (MC) simulation is considered as the most accurate method. A number of software packages have been developed in the past for the MC simulation of radiation transport, ranging from those packages for general purposes, such as EGS4 (Nelson *et al.*, 1985; Bielajew *et al.*, 1994), EGSnrc (Kawrakow, 2000), MCNP (Briesmeister, 1993), PENELOPE (Baró *et al.*, 1995; Salvat *et al.*, 1996; Sempau *et al.*, 1997; Salvat *et al.*, 2009), GEANT4 (Agostinelli *et al.*, 2003), to those clinically oriented ones such as VMC++ (Kawrakow *et al.*, 1996), MCDOSE/MCSIM (Ma *et al.*, 1999; Li *et al.*, 2000; Ma *et al.*, 2002), and DPM (Sempau *et al.*, 2000), to name a few.

Since MC simulation is a statistical method, its accuracy largely depends on the number of simulated particle histories. In a MC simulation, we compute, from first principles, how a particle evolves step by step. A large number of histories are simulated in order to achieve a desired statistical accuracy. Therefore, despite the vast development in computer architecture and increase of processor clock speed in recent years, the efficiency of the currently available full MC dose engines is still not completely satisfactory for routine clinical applications in radiotherapy treatment planning.

One way out of this obstacle is to perform the computation in a parallel fashion by taking advantages of advanced computer architectures, such as CPU clusters or general purpose graphics processing units (GPUs). Compared to CPU clusters of similar parallel computing power, GPU is easier to access and maintain (it can run on a local desktop computer) and much less expensive (one to two orders of magnitude lower cost). With affordable graphic cards such as NVIDIA’s GeForce, GTX, and Tesla series, GPU-based computing has recently been utilized to speed up heavy duty computational tasks in radiotherapy, such as cone-beam CT reconstruction (Xu and Mueller, 2005; Li *et al.*, 2007; Sharp *et al.*, 2007; Xu and Mueller, 2007; Yan *et al.*, 2008; Jia *et al.*, 2010), deformable image registration (Sharp *et al.*, 2007; Samant *et al.*, 2008; Gu *et al.*, 2010), dose calculation (Jacques *et al.*, 2008; Hissoiny *et al.*, 2009; Gu *et al.*, 2009), and treatment plan optimization (Men *et al.*, 2009). In particular, Jacques *et al.* (2008) and Hissoiny *et al.* (2009) have explored GPUs for fast dose computation via the superposition/convolution algorithm and Gu *et al.*(2009) implemented a finite size pencil beam model for GPU based dose calculation.

In this paper, we report our recent development of a Monte Carlo dose calculation package under the Compute Unified Device Architecture (CUDA) platform developed by NVIDIA (NVIDIA, 2009), which enables us to extend C language to program GPU. At University of California San Diego (UCSD), we have implemented the full MC dose calculation package DPM with coupled electron-photon transport based on reasonably priced and readily available GPUs. Our code is benchmarked against the original sequential DPM package on CPU. The roadmap of this paper is as follows. In Section 2, we describe the general structure of the DPM package and some key issues of our implementation. Section 3 presents experimental results of our dose calculation in heterogeneous phantoms. The computational time, as well as simulation results, is compared between GPU and CPU implementations. Finally, we conclude our paper in Section 4 and discuss possible limitations of our GPU implementation.

## 2. Methods and Materials

### *2.1 DPM MC simulation*

The Dose Planning Method (DPM) Monte Carlo code has been developed for dose calculations in radiotherapy treatment planning (Sempau *et al.*, 2000). It employs an accurate and efficient coupled electron–photon transport scheme and a set of significant innovations for transporting particles in order to gain performance. The electron transport part of DPM relies on a condensed history technique for elastic collisions. Detailed (*i.e.*, step-by-step) simulation is used for inelastic collisions and bremsstrahlung emission involving energy losses above certain cutoffs. Below these cutoffs, the continuous slowing down approximation (CSDA) is employed. As for the photon transport, DPM implements the Woodcock tracking method, which significantly increases the simulation efficiency of the boundary tracking process (Woodcock *et al.*, 1965). DPM also focuses on a small dynamic range (in energy and material) of radiotherapy problems, where some approximations are valid, so as to speed up the calculation. The accuracy of DPM has been demonstrated to be within $\pm 2\%$ for both clinical photon and electron beams (Chetty *et al.*, 2002; Chetty *et al.*, 2003).

### *2.2 CUDA implementation*

Our GPU-based DPM code is developed using CUDA as programming environment and NVIDIA GPU cards as hardware platform. Our implementation is an exact translation of the original sequential DPM code into a parallel version on GPU. Specifically, we treat each computational thread on GPU as an independent computing unit, which tracks complete histories of a source particle as well as all secondary particles it generates, as if in the original sequential DPM code. This multiple-thread simulation is performed in a batched fashion. A large number of CUDA threads run simultaneously in every batch and efficiency can be gained due to this vast parallelization.

This computation procedure is schematically depicted in Figure 1. After initializing the program, we first load all the relevant data from files on the hard disk to GPU global memory. For example, cross section data are prepared as a function of energy with respect to all physical interactions and materials. Random seeds are generated on CPU and then passed to GPU in this step as well. After this preparation stage, a special C function, termed *kernel*, is invoked by CPU and is executed $N$ times on a total number of $N$ threads in a parallel manner on GPU, where each thread performs the calculation independently, as illustrated by the big dash box in Figure 1. There are three steps within each thread. a) Local counter initialization. A local counter, an array privately owned by a thread for scoring the dose deposition from the particle histories followed by the thread, is initialized. b) Particle simulation. In this step, an empty stack is first initialized. A source particle is generated according to a source model. Simulation with respect to this particle is then performed and dose information is stored on the local counter whenever dose deposition occurs. In this process, any secondary particles due to some physical interactions, *e. g.* Compton scattering, are recorded in the stack. Upon completing the simulation for the current particle, another particle is fetched from the stack and simulation with respect to the particle is performed. Such a process is repeated till the stack becomes empty. c) Dose collection. The dose deposition obtained from each thread is transferred onto a global dose counter and summation of the dose deposition is performed where necessary. Such a kernel is executed repeatedly till a preset number of history is achieved. Finally, the dose on the global counter is transferred from GPU memory to CPU memory and output before the entire program terminates.

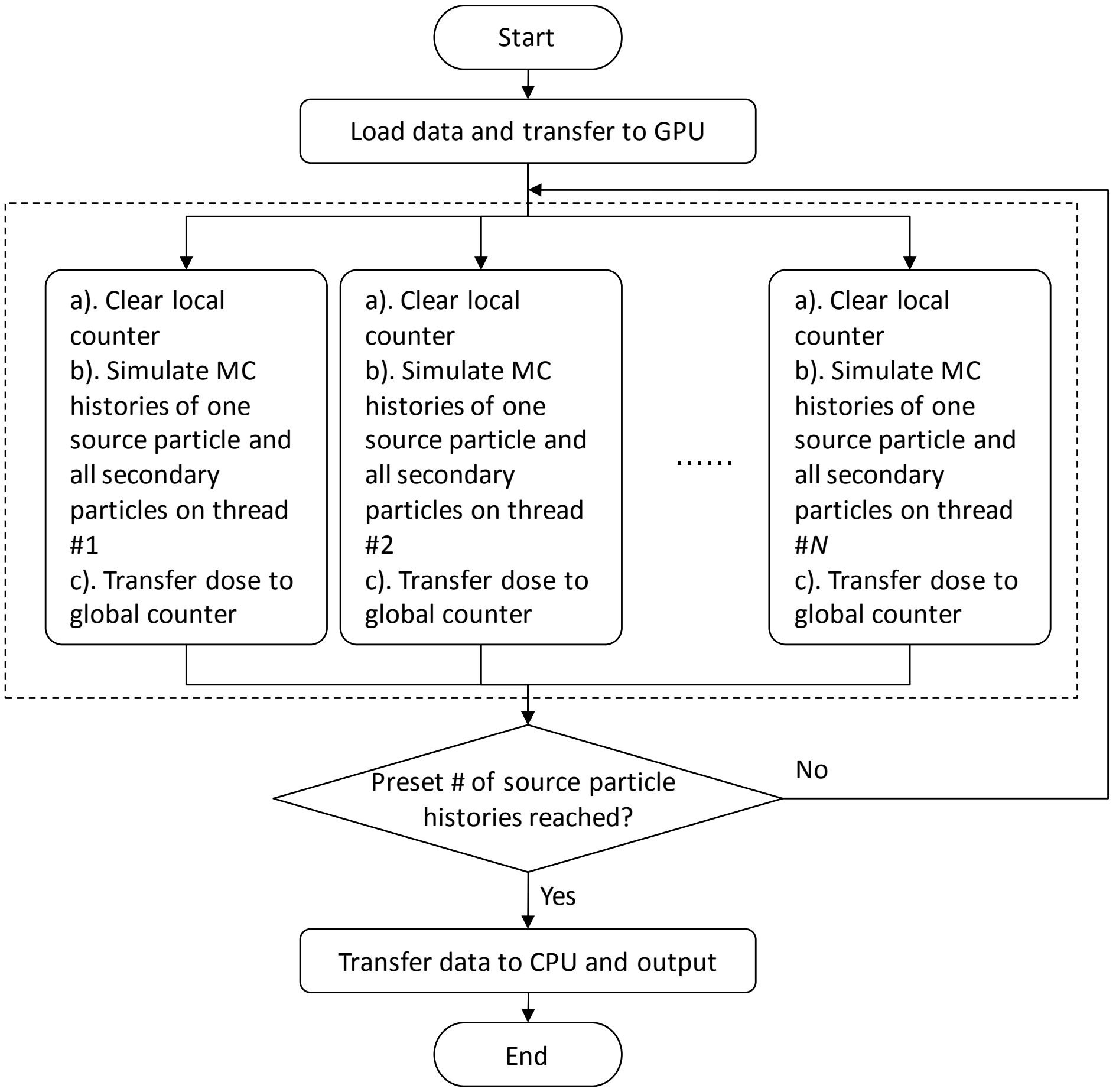

**Figure 1**. The flow chart of our GPU-based DPM Monte Carlo simulation. The area bounded by dash lines indicates the kernel simulating histories of one source particle as well as all secondary particles. The kernel is executed on GPU *N* times with *N* different CUDA threads in parallel.

*2.3 Data structure*

Data structure on GPU is an important issue in the MC simulation. Since double-precision arithmetic is not fully supported in CUDA, throughout our implementation, single-precision floating point data type is used to represent rational numbers instead of double-precision as in the original sequential DPM code. This, however, seems not reduce the simulation accuracy, as will be seen in the results presented below.

In our simulation, variables are stored differently in GPU memory depending on how they are accessed. First, read-only data are those accessed by each thread frequently during the simulation without modification. It is desirable to store data of this type on the constant memory of GPU, which provides a high access speed. However, only those variables of small sizes are stored in this way due to the limitation of space in constant memory on GPU (~64k). As for those read-only data of large sizes, we store them on GPU global memory in the form of texture memory. Since accessing texture memory is cached, high performance speed can be achieved during the simulation. Examples in this category are the cross section data for all types of collisions as a function of energy in each material. Finally, those data constantly being modified are stored on the global memory of GPU. For instance, it is necessary for each thread to keep and update its own random number seeds in order to generate independent random number sequences. The local dose counter mentioned above also belongs to this category, which is frequently modified to record the dose deposition for the particle histories tracked by the particular thread only. The use of this data structure, however, has two limitations on the performance of our code. In the first place, there is a huge demand of GPU memory in order to store the variables of this kind, since most of them have to be allocated multiple times, one for each thread, during the simulation. As a result, the number of threads is restricted by the total amount of GPU memory, which limits the capability of the parallelization. Second, global memory is accessed in an un-cached way and thus using it is very time-consuming.

*2.4 Other technical issues*

One technical detail worth addressing here is how the dose deposition is recorded in our simulation. Within the MC simulation kernel, each thread updates its own local counter every time energy is deposited in a voxel. Since it is not possible to predict where a dose deposition will occur, the simplest approach seems to establish a one-to-one correspondence between voxels and the elements of the local counter array, so that each element keeps tracking the energy deposited in the corresponding voxel. Nonetheless, neither is this strategy memory efficient nor necessary. Indeed, the energy is only deposited to a few number of voxels within histories of a source particle and all subsequent particles it generates, as these particles only travels through a small fraction of total voxels. It is therefore feasible to allocate the local dose counter with a predetermined length, which records the amount of energy being deposited and the location where the corresponding deposition occurs. The length of this local counter has to be carefully chosen to be long enough.

Upon exiting the kernel, all threads transfer dose deposition from their own local counters to the global counter. In this process, energy deposited to a particular voxel is summed over different threads. This, however, potentially causes a memory conflict due to the attempt of writing to a same memory address from different threads. We adopt an atomic float addition function developed by Lensch *et al.* (Lensch and Strzodka, 2008) to resolve this problem. This function is atomic in the sense that, once one thread is writing to a memory address during the summation, no other threads can interference this process.

Another key issue in MC simulation is the random number generator, whose quality plays an important role in the simulation. In our implementation, we take the pseudo-random number generator RANECU (L'Ecuyer, 1988), which has a total period of ~ $2\times10^{18}$. In addition, different CUDA threads are initialized with different random seeds generated by a CPU random number generator. During the simulation, each CUDA thread deals with its own random seeds, so that the random number sequences produced from distinct threads are statistically independent.

### *2.5* GPU *card*

Computer graphic cards, such as the NVIDIA GeForce series and the GTX series, are conventionally used for display purpose on desktop computers. It typically consists of 32-240 scalar processor units and 256 MB to 1 GB memory. Recently, NVIDIA introduced special GPUs solely dedicated for scientific computing, for example the Tesla C1060 card that is used in our MC simulation. Such a GPU card has a total number of 240 processor cores (grouped into 30 multiprocessors with 8 cores each), each with a clock speed of 1.3 GHz. The card is equipped with 4 GB DDR3 memory, shared by all processor cores.

## 3. Experimental Results

We investigated the performance of our GPU-based DPM MC simulation on various phantom experiments. Two testing phantoms are considered in our simulation, both of which are $30.5\times30.5\times30.0\ \text{cm}^3$ rectangular cuboids with a voxel size of $0.5\times0.5\times0.2\ \text{cm}^3$. Along the $z$ direction, the phantoms consist of three layers, namely either water-bone-water layers of thicknesses $5\ \text{cm}$, $5\ \text{cm}$ and $20\ \text{cm}$, respectively, or water-lung-water layers of same dimensions. Either an electron or a photon point beam sources with $\text{SSD} = 90\ \text{cm}$ impinges normally on the center of the phantoms at the *x*-*o*-*y* plane. The electron beam is chosen to be 20 MeV mono-energetic, while the photon beam is generated according to a realistic 6 MV energy spectrum. For both the electron and the photon beams, field sizes are set to $10\times10\ \text{cm}^2$ at the isocenter with $\text{SAD} = 100\ \text{cm}$. The absorption energies are 200 keV for electron and 50 keV for photon. A total number of $10^7$ and $10^9$ particle histories are simulated for the electron source and the photon source cases, respectively. The statistical uncertainty of our simulation is characterized by the averaged relative uncertainties $\overline{\sigma_D/D}$, where $\sigma_D$ is the standard deviation of the local dose $D$ at a voxel. The over bar stands for an average over the region where the local dose $D$ exceeds half of its maximum value in the entire phantom. The number of particle histories simulated is large enough, so that $\overline{\sigma_D/D}$ were found to be less than 1% in all cases studied. In order to validate our GPU code and test its efficiency, we also run the original sequential DPM code on CPU and used the results as reference for comparison. The original DPM code is executed on a 2.27 GHz Intel Xeon processor, while the GPU code is on a NVIDIA Telsa C1060 card.

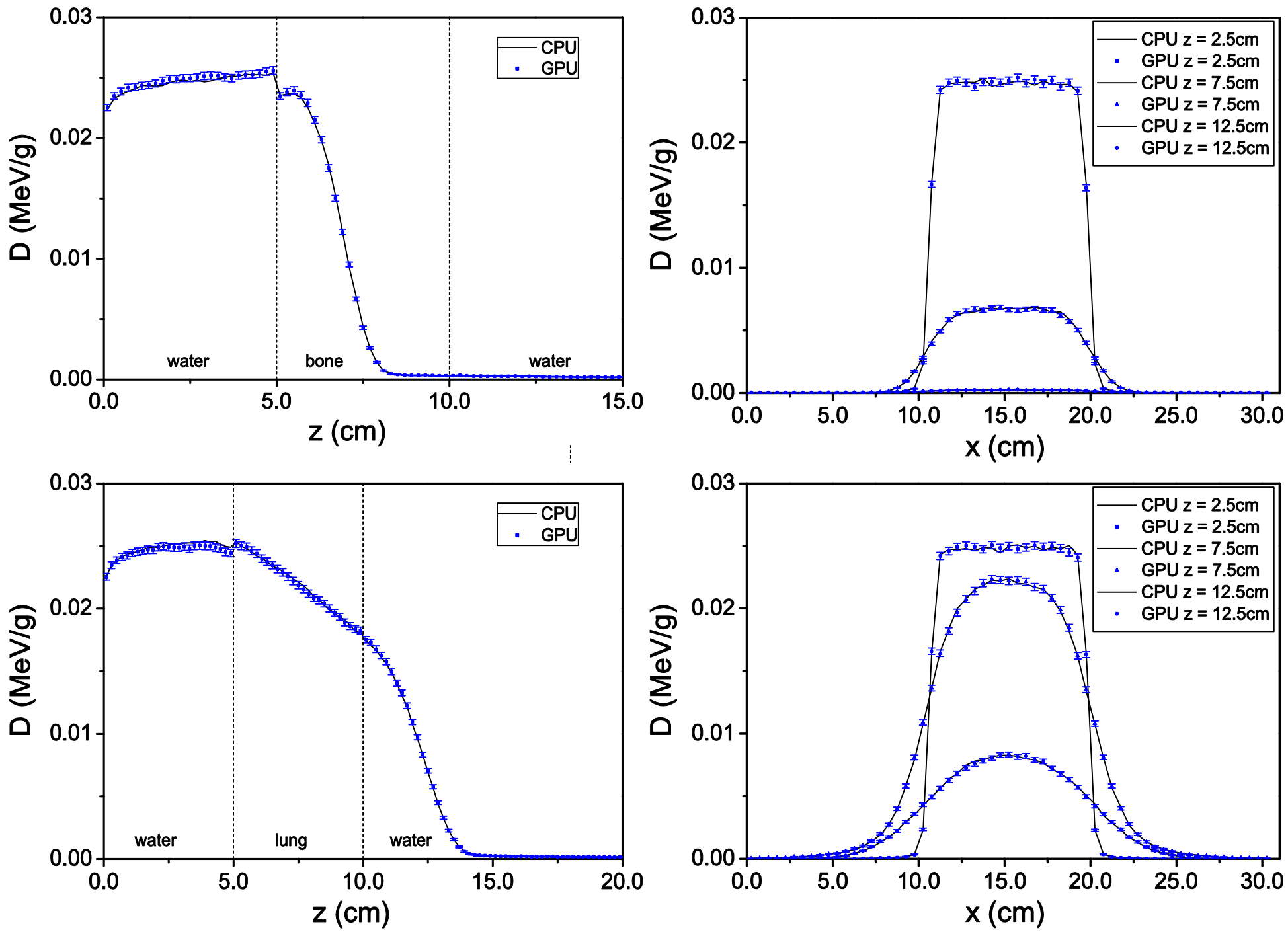

**Figure 2**. Depth-dose curves (left column) and lateral dose profiles at different depths (right column) of a 10×10 cm$^2$, 20 MeV electron point source at $SSD$ = 90 cm impinging on a water-bone-water phantom (top row) and a water-lung-water phantom(bottom row). Error bars correspond to 2 standard deviations.

Figure 2 and Figure 3 demonstrate the dose distribution for the tested electron and photon beam cases, respectively. In both figures, the left columns are the depth dose curves along the central axis, while the right columns correspond to the lateral dose profiles at different depths. The simulation results for the water-bone-water phantom and the water-lung-water phantom are shown in the top rows and the bottom rows, respectively. The error bars represent $2\sigma_D$ levels of the results. For the purpose of clarity, we did not draw error bars for the results obtained from the CPU code, which are of similar sizes to those for the results from GPU. Results obtained from CPU and GPU are found to be in good agreement with each other. The difference at more than 98% of the calculation points is within 1% of the maximum dose all testing cases. In Table 1, we list the average relative uncertainties $\overline{\sigma_D/D}$ in all testing cases, which are controlled to be less than 1% in our calculation.

Table 1 also depicts the computation time in the two testing cases. $T_{CPU}$ stands for the execution time of the CPU implementation, while $T_{GPU}$ is that of the GPU implementation. In all time measurements, the time $T$ correspond to the complete MC simulation time, including the time used for program initialization, finalization, and data transferring from CPU to GPU in the GPU implementation. Speed-up factors of about 5.0 ~ 6.6 times have been observed in the GPU calculation compared to CPU simulation. These speed-up factors on GPU are achieved using 100 CUDA blocks and 128 threads per block. These numbers regarding GPU configurations are carefully tuned in order to achieve the best performance.

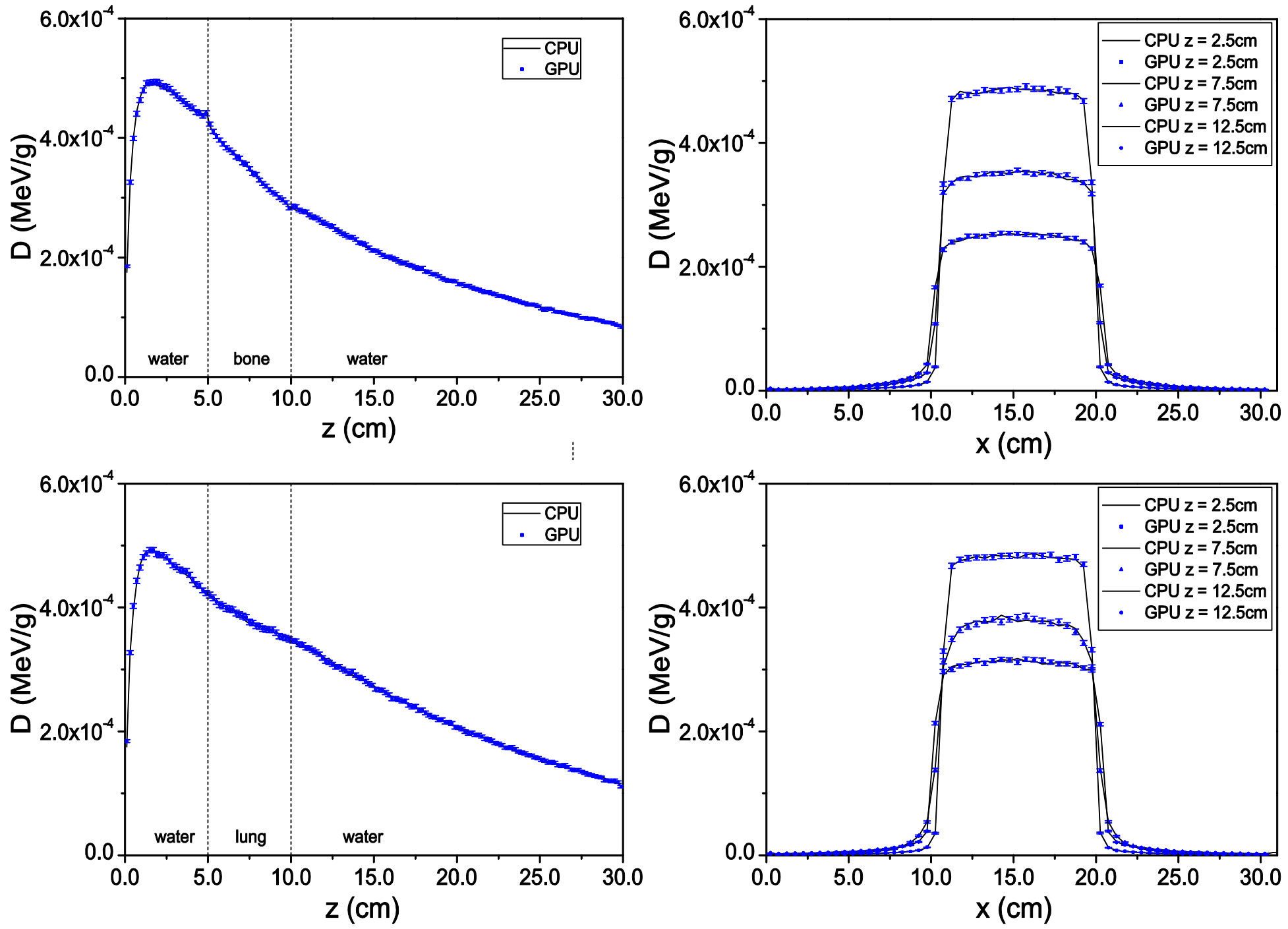

**Figure 3**. Depth-dose curves (left column) and lateral dose profiles at different depths (right column) of a 10×10 cm$^2$, 6 MV photon point source at SSD = 90 cm impinging on a water-bone-water phantom (top row) and a water-lung-water phantom(bottom row). Error bars correspond to 2 standard deviations.

**Table 1**. Average relative uncertainty ($\sigma_D/D$) execution time ($T$), and speedup factors ($T_{CPU}/T_{GPU}$) for four different testing cases.

| Source type | # of Histories | Phantom | $\overline{\sigma_D/D}$ CPU (%) | $\overline{\sigma_D/D}$ GPU (%) | $T_{CPU}$ (sec) | $T_{GPU}$ (sec) | $T_{CPU}/T_{GPU}$ |
|---|---|---|---|---|---|---|---|
| 20MeV Electron | $10^7$ | water-lung-water | 0.65 | 0.65 | 470.0 | 80.6 | 5.83 |
| 20MeV Electron | $10^7$ | water-bone-water | 0.63 | 0.63 | 508.0 | 101.8 | 4.99 |
| 6MV Photon | $10^9$ | water-lung-water | 0.51 | 0.52 | 5615.0 | 845.8 | 6.63 |
| 6MV Photon | $10^9$ | water-bone-water | 0.45 | 0.46 | 6964.0 | 1242.3 | 5.61 |

## 4. Discussion and Conclusions

In this paper, we have successfully implemented the DPM Monte Carlo dose calculation package on GPU architecture under NVIDIA CUDA platform. We have also tested the efficiency and accuracy of our GPU implementation with respect to the original sequential DPM code on CPU in various testing cases. Our results demonstrate the adequate accuracy of our implementation for both electron and photon sources. Speed-up factors of about 5.0 ~ 6.6 times have been observed. The code is in public domain and available to readers on request.

MC simulations are known as embarrassingly parallel because they are readily adaptable for parallel computing. It has been reported that the DPM package has been parallelized on a CPU cluster, and roughly linear increase of speed has been realized with respect to an increasing number of processors when up to 32 nodes of an Intel cluster are involved (Tyagi *et al.*, 2004). Nonetheless, this liner scalability is hard to achieve on GPU architectures, when simply distributing particles to all threads and treating them as if they were independent computational units.

In general, the means of performing parallel computation are categorized into *Task Parallelization* and *Data Parallelization*. MC simulation, a typical task parallelization problem, is preferable for a CPU cluster developed through, for example, *message passing interface* (MPI). All particles histories simulated in an MC dose calculation can be distributed to all processors, which execute simultaneously without interfering with each other. Only at the end of the computation will the dose distribution need to be collected from all processors. Apparently the parallelization of this manner is capable of speeding up the simulation easily with a large number of CPU nodes. On the other hand, GPU is known as suitable for the Data Parallelization problems. A GPU multiprocessor employs an architecture called *SIMT* (single-instruction, multiple-thread) (NVIDIA, 2009). Under such architecture, the multiprocessor executes program in groups of 32 parallel threads termed *warps.* If the paths for threads within a warp diverge due to, *e.g.*, some *if-else* statements, the warp serially executes one thread at a time while putting all other threads in an idle state. Thus, high computation efficiency is only achieved when 32 threads in a warp process together along the same execution path. Unfortunately, in a MC calculation the computational work paths on different threads are statistically independent, essentially resulting in a serial execution within a warp. Since there are physically only 30 multiprocessors on a GPU, our simulation is indeed parallelized by just 30 independent computation units. Considering, furthermore, the GPU clock speed is 1.3 *GHz*, about a half of that for CPU, the highest possible speed-up factor for a GPU-based MC simulation will be roughly 15 times.

Other factors may also adversely restrict our simulation efficiency, such as the memory access pattern. Since all threads share the usage of a global memory in our code, the random access to different memory addresses from different threads produces a serious overhead. In addition, the global memory on GPU is not cached, leading to 400-600 clock cycles of memory latency, while the CPU memory is normally cached and favorable for fast fetching. In a nut shell, due to many factors limiting our simulation efficiency, our code has only achieved speed-up about 5.0~6.6 times.

On the other hand, the GPU implementation of an MC simulation still has the obvious advantage at its low cost and easy accessibility as opposed to the CPU clusters. Our work has clearly demonstrated the potential possibility to speed up an MC simulation with the aid of GPU. Currently, another MC simulation algorithm which is specifically tailored for GPU architecture is under development and a boost of MC simulation speed is expected.

## Acknowledgements

This work is supported in part by the University of California Lab Fees Research Program. We would like to thank NVIDIA for providing GPU cards for this project. JS acknowledges partial financial support from the Spanish Ministerio de Educacion y Ciencia, project no. FIS2006-07016.